\documentclass[12pt]{extarticle}
\usepackage[utf8]{inputenc}
\usepackage[T1]{fontenc}
\usepackage{amsmath,amssymb}
\usepackage{enumitem}
\usepackage{float}
\usepackage{adjustbox}
\usepackage{graphicx}
\usepackage{booktabs}
\restylefloat{table}

\usepackage{cite}
\usepackage[linktocpage=true, colorlinks=false, linkcolor = blue, citecolor = red]{hyperref}
\begin{document}
\title{Universal Constants \\ as Manifestations of Relativity}

\author{A. A. Sheykin\thanks{\texttt{anton.shejkin@gmail.com}}\\
\textit{Saint Petersburg State University, St. Petersburg, Russia}}
\date{}
\maketitle
\abstract{We study the possible interpretation of the "universal constants" by the classification of J.-M.~L\'evy-Leblond. $\hbar$ and $c$ are the most common examples of constants of this type. Using Fock's principle of the relativity w.r.t. observation means, we show that both $c$ and $\hbar$ can be viewed as manifestations of certain relativity. We also show that there is a possibility to interpret the Boltzmann's constant in a similar way, and make some comments about the relativistic interpretation of the constant spacetime curvature and gravitational constant $G$. 
}

	\section{Introduction}
	Since the beginning of XX century, theoretical physics has been faced many challenges related to the interpretation of its results. These challenges made obvious the need for deep understanding of the methodology of physics. The two main branches of contemporary physics in which this need is especially sharp are undoubtedly the theory of relativity, whose appearance led to long and heated debates about the meaning of relativity, and the quantum theory, whose interpretation remains the subject of discussions. Methodological discussions can result in the simplification and elucidation of the basic principles of the theory. A flashing example of this simplification can be seen in the case of Lorentz transformations, which can be derived in many different ways, and some of these ways are certainly more preferred than the others from a contemporary point of view. However, it must be stressed that these discussions can also stimulate the development of physical theories: for example, the Einstein's dissatisfaction of the methodology of quantum mechanics (QM) led him to propose the EPR paradox which turned out to be the starting point of the theory of quantum entanglement. 
	
	The vast majority of physicists surely recognized the importance of methodology of science, and it is hard to find a person who, being a significant figure in a particular field of physics, did not contribute to its methodology. Vladimir Fock is not an exception: as the other creators of QM (such as Schr\"odinger, Born and Ehrenfest, with whom he corresponded), he was deeply involved in methodological discussions. This involvement started shortly after the appearance of the abovementioned EPR paper, to which Fock replied by the proposition of the \textit{principle of relativity with respect to observational means}. According to Fock, one can resolve the issue of the apparent incompleteness of quantum mechanics via this principle. In his late paper Fock, however, states that this principle could have a wider area of applicability. The main aim of this paper is to follow this hypothesis and to apply this principle to a broader class of theories and to point out that there are theories to which this principle cannot be applied (Fock was a well-known opponent of the concept of general relativity in the Einstein's interpretation).
	
	We identify the class of theories, to which this principle can be applied, using the ideas of another prominent physicist, who made significant contributions to the methodology of physics, namely J.-M.~L\'evy-Leblond. As V.~A.~Fock, he started his career mainly in the field of quantum physics. His renowned papers on the group theory in QM were followed by methodological discussions of the special relativity and interpretations of QM and well of the roles of constants in various physical theories. 
	
	In particular, we will focus our attention on the concept of \textit{universal constant}, which appeared in the classification scheme of fundamental constants developed by L\'evy-Leblond. Our goal is to show that each constant that belong to this type points to the presence of relativity of particular type, and vice versa: if the constant is not universal, the corresponding relativity is not the relativity with respect to the observational means.
	
	This paper is organized as follows. In Section \ref{Fock}, we review Fock's development of the principle of relativity w.r.t. observational means. In Section \ref{const}, we, following L\'evy-Leblond, introduce the concept of universal constant, identify ($c$ and $\hbar$) as belonging to this class and show that they can be interpreted as a manifestation of the relativity w.r.t. observational means. In Section \ref{rel} we discuss the possibility to interpret the Boltzmann constant $k_{\text{B}}$ in terms of this relativity and make some comments about possible interpretation of the constant spacetime curvature $R$ \cite{2005.08196}. In Section \ref{GR}, we discuss the status of the gravitational constant $G$ and the notion of general relativity. 
	
	\section{Principle of Relativity with respect to Observational Means}\label{Fock}
	As it was mentioned above, in 1936 Einstein, Podolski and Rosen raised the question about the incompleteness of quantum description of reality\cite{EPR}. In the introductory note that Fock wrote in the supplement to the Russian translation \cite{Fock1936} of the EPR paper, Fock states that wavefunction, in contrary with Einstein's opinion, resembles not an "objective state of the system", but rather a result of a "maximally precise experiment". It must be stressed that this appellation to the unbreakable connection between the description of the quantum system and the experimental setup will become the cornerstone of Fock's future works on the interpretation of quantum mechanics. He then proposes a resolution of the EPR paradox using the fact that we do not necessarily know a result of such an experiment for a generic system and, therefore, unable to assign a wavefunction to it. Such systems must be described by a statistical operator (i.e., density matrix). 
	
	The ideological discussions about the quantum theory in the USSR (see \cite{graham}, ch. 10) forced Fock to write several explanatory papers about its structure and interpretation . He strengthened his earlier statement by the observation that the quantum description of reality is not absolute  \cite{Fock1938}, and the elementary entity that has to be considered in the quantum theory is not merely a quantum system, but rather a quantum system interacting with some measuring device \cite{Fock1949}.
	
	At the same time, a similar discussion occurred between leading Western scientists; the leading role in it was played by Niels Bohr \cite{Bohr1948}. The disagreement with some of the Bohr's views on the QM \cite{Fock1955} led Fock to the long debate with him (a detailed comparison of the epistemological positions of Fock and Bohr can be found in \cite{martinez2019}). As a result of this debate, the paper \cite{Fock1957} appeared, in which Fock for the first time explicitly stated the following:
	\begin{quote}
		Not only accuracy in the quantitative sense, but also formulation of qualitatively new
		properties of micro-objects requires new methods of description, and above all, a new element of relativity --- relativity with respect to the observation means --- has to be introduced.
	\end{quote}
	Fock stresses that this relativity does not imply the absence of an objective description of quantum systems:
	\begin{quote}
		Even in classical physics such simple concepts as the trajectory of a mass point, being wholly objective, are at the same time relative, because they have a definite meaning
		only in a definite frame of reference. Similarly, in quantum physics relativity
		to the means of observation only renders physical concepts more exact and
		allows the introduction of new concepts, but in no case does it deprive them of
		their objectivity.
	\end{quote}

	The relativity w.r.t. observational means differs from relativity w.r.t. a frame of reference by the fact that the latter can be formulated as some continuous symmetry of equations of motion (or action of theory, if it has a Lagrangian or Hamiltonian formulation), whereas the former puts some non-dynamic restrictions on the variables of the theory. In quantum mechanics these restrictions are represented by Heisenberg uncertainty relations:
	\begin{align}\label{Heis}
		\Delta p \Delta x \geq \frac{\hbar}{2}, \ \Delta E \Delta t \geq \frac{\hbar}{2}, 
	\end{align}
	and so on. Fock, however, suggested that these relations should not be interpreted as manifestations of some uncertainty, but rather as manifestations of the limited applicability of classical models (and therefore manifestations of relativity)\cite{Fock1949}.  
	
	This is the essence of the principle of relativity w.r.t. observational means, which we will call PeRelOM throughout this paper. {Russian \emph{perelom} means "a pivotal moment" or "a turning point", since, in our opinion, the discovery of this principle indeed could become a pivotal moment in physics. Copernicus proclaimed the relativity of position (\cite{bondi1968}, p. 13), Galilei --- the relativity of motion (\cite{yaglom1979}, p. 17), Mach --- the relativity of inertia (\cite{bondi1968}, p.29), Einstein --- the relativity of simultaneity (\cite{einstein1920}, ch. IX), and Fock (using Heisenberg's concepts) --- the relativity of observables.} 
	
	For Fock, the main feature of PeRelOM was probably the fact that it addresses Einstein's objections about the incompleteness of QM. According to Fock, QM is no more and no less complete than the classical theory, if one takes into account the fact that the majority of characteristics of classical systems are relative w.r.t. a frame of reference. However, it does not provide the solution (and even hints of it) of some other problems of QM, e.g., related to the reduction of the wavepacket. Since the main part of works on the interpretation of the QM deals with exactly these problems, PeRelOM did not draw much attention of the quantum-mechanical community. It should be noted  though that similar approaches  were later proposed by several authors independently of Fock. For example, accordind to the authors of \cite{chiatti2014}, there is a connection between PeRelOM and transactional interpretation that was proposed by Kramer in the 1986. The rediscovery of the relational nature of QM by Rovelli \cite{quant-ph/9609002} is also worth mentioning.

	In the later years, Fock used PeRelOM throughout his papers on the interpretation of QM, refining its formulation (the most detailed one is given in \cite{Fock1971}) and suggesting some other areas of its applicability. For example, in \cite{Fock1971} he noticed the similarity between the pairs "quantum system --- observer" and "living creature --- environment" and conjectured that PeRelOM could be applied to the analysis of biologic systems. In another paper \cite{Fock1973}, Fock made some comments about the possible application of PeRelOM to Einsteinian gravity (we will discuss this question in Section \ref{GR}). In order to extend PeRelOM beyond quantum mechanics, in the next section we will try to identify the areas of physics where PeRelOM can (or can not) be established using the notion of universal constant.

	\section{Classification of fundamental constants}\label{const}
	The question of the conceptual nature of physical constants has been widely discussed for more than a century. There are numerous attempts to explain their appearance and determine their role in the laws of nature (see, e.g., \cite{barrow2003constants}. Many references can also be found in \cite{2005.08196}). Among these attempts, the one that is especially useful in the context of this paper was proposed by J.-M.~L\'evy-Leblond. According to \cite{LL1977}, the full set of constants can be split into three parts:
	\begin{enumerate}[label={\Alph*}.]
		\item Characteristics of the objects,
		\item Characteristics of the phenomena, 
		\item \emph{Universal constants}. 
	\end{enumerate} 
	In the field of elementary particle physics, the masses of particles as well as their quantum numbers belong to type A. The constants of interaction ($G$, $e$, $g_s$ and $g_w$) belong to type B. The remaining constants of particle physics that cannot be related to the specific particle or interaction are $c$ and $\hbar$. L\'evy-Leblond characterizes them as universal constants. The role of these constants, according to L\'evy-Leblond, is to synthesize concepts (or even theories). Namely, the universal constant $c$ forms a new geometric object (spacetime) out of pre-existing concepts of space and time, making them parts of the entirely new concept that was absent in any of the previous theories describing one or another. The Planck constant $\hbar$, in turn, forms a concept of a quantum object, or \emph{partiqle} \cite{LL1981}, out of concepts of particle and wave. L\'evy-Leblond stresses that quantum objects are neither particles nor waves, although in the specific conditions their observed characteristics might match the characteristics of the one or another.
	
	L\'evy-Leblond described the primary aim of these characteristics as the consideration of the evolution of the conceptual role of physical constants. The particular type is not permanently assigned to some constants, but rather characterizes its role in physical theory at the specific moment. For example, the speed of light $c$, which was initially introduced as type A constant, was then promoted to type B by Maxwell, who showed that it characterizes the entire electromagnetic interaction, and then to the type C by Einstein who revealed its kinematic nature (geometrized a few years later by Minkowski).  
	
	Besides modern universal constants $c$ and $\hbar$, there are other constants of type C, which L\'evy-Leblond calls classical ones. These constants also served as synthesizers of certain concepts, but these concepts had entirely lost all their particular significance, and the corresponding constant vanishes, since in this theory there is nothing to synthesize anymore. To this class belongs the Boltzmann constant $k$, which unifies a temperature and kinetic energy; and mechanical equivalent of heat $J$, which shows that heat can be transformed to work and vice versa. While we completely agree with the proposition that $J$ nowadays vanishes from physics, we can not wholly agree with the analogous point of view on $k$. We will discuss its role and fate in the Section~\ref{rel}.

	\section{Universal constants and relativity}\label{rel}
	Now we can put together the ideas of Fock and L\'evy-Leblond in the following way. First of all, let us notice that the universal constant $\hbar$ appears in the Heisenberg uncertainty relation, which, in turn, can be interpreted as a sign of the relativity w.r.t. observational means. The immediate reaction to this connection is a curious question: are there other uncertainty relations that contain universal constants?

	The answer is surely positive. Let us consider a famous example from the signal processing theory. Suppose that we have some signal $\phi(t)$, where $x$ is a time parameter and $\phi$ is an amplitude of the signal. We can also define a Fourier transform of the signal:
	\begin{equation}
	\phi(f) = \int \limits_{-\infty}^{+\infty} dt \phi(t) e^{ift},
	\end{equation} 
	where $f$ is a frequency of the signal. There is an expression known as Gabor limit \cite{gabor1946} that connects uncertainties in transmission time and frequency band of the signal:
	\begin{equation}\label{gabor}
	\Delta t \Delta f \geq 1/2.
	\end{equation}
	It means that one can not localize the signal in time and frequency at once: the best one can achieve is a Gaussian profile of the signal in $t$ and $f$ variables (Hardy's theorem). Note that in this form, it is a purely mathematical fact which is a consequence of the properties of Fourier transform  (see, e.g. \cite{Grochenig2001}). 
	
	To give it a physical interpretation, let us take into account the interaction of the signal with the measuring device, i.e., receiver. Suppose that the signal is a wave of some kind, which is characterized by wavelength $\lambda$ and constant\footnote{If the speed of the wave can vary (e.g., in a dispersive medium), the non-linearity could appear in the system, which would make difficult its Fourier analysis since Fourier transform is a linear operation.} speed $c$, so
	\begin{equation}\label{speed}
	f=c/\lambda.
	\end{equation}
	Putting it into \eqref{gabor} and dividing all by $c$, we obtain
	\begin{equation}\label{gaborc}
	\Delta t \Delta \left(\frac{1}{\lambda}\right) \geq \frac{1}{2c}.
	\end{equation}
	Now let us consider some specific experimental setup. Suppose that our receiver (whose characteristic size is comparable to characteristic wavelengths of the signal) interacts with some signal, which is strictly localized in time. According to \eqref{gabor}, we must face a huge uncertainty in frequency. However, \eqref{speed} shows that in case of large $c$, this huge uncertainty corresponds to modest uncertainty in wavelength, so the signal can be \emph{effectively} localized both in time and in wavelength. Reversely, if the signal has a thin frequency band, it takes a long time to receive it in its completeness (it is often said that such signal has "long tails"), but if the signal propagates very fast, these very intervals of time are small in comparison to $c/\lambda$.
	
	As a result, we can see that the constant $c$ in the Gabor limit modified as above \eqref{gaborc} plays exactly the same role that Planck constant $\hbar$ plays in the Heisenberg uncertainty relation \eqref{Heis}. Namely, it connects the properties of the considered phenomena with our observational abilities. In this sense, inequality \eqref{gaborc} has a "classical limit" as well as   \eqref{Heis}.  Indeed, when the action of a physical system is sufficiently large, the simultaneous measurement of its position and momentum becomes possible. Analogously, when the signal propagation speed is sufficiently large, simultaneous measurement of its transmission time and wavelength becomes possible. 
	
	It must be stressed that the existence of this limit is connected with properties of the observational means: one can reach this limit in the above examples, if $\Delta t \ll L/c$, where $L$ is a characteristic scale of a receiver (e.g. if this is a so-called "wave antenna" or monopole antenna), or $T \Delta(1/\lambda) \ll 1/c$,  where  $T$ is a period of free oscillations that can exist in it (if the antenna itself is not \emph{self-resonant}).
	
Another important fact is that the presence of $c$ in the above example is not, strictly speaking, necessarily related to special relativity (if the latter is considered in the context of relativity w.r.t reference frame). The constant $c$ connects specific quantities without a direct indication of any symmetry of the system (except its linearity). For example, it can be associated with the speed of sound, in which case the resulting uncertainty principle is by no means fundamental (see, e.g. \cite{1208.4611}). However, the identification of $c$ with the speed of light allows one to connect this constant with an unavoidable (at least in principle) relativity w.r.t. observational means, in the same way as $\hbar$. 
	
	It is also worth noting that alongside with $c$, there is another fundamental constant of geometric (or, so to say, kinematical) origin: namely, the constant spacetime curvature $R$. Its role in kinematic was discovered and discussed by L\'evy-Leblond and Bacry \cite{LL1977}, for a further discussion and references, see  \cite{2005.08196}. Analogously to $\hbar$ and $c$, it also can serve as a concept synthesizer, connecting notions of length and angle. This synthesis can be seen at the level of commutation relations of the spacetime symmetry group generators, one of whose has the following form:
	\begin{equation}
	[P_i,P_j] = \mp\frac{1}{R^2} M_{ij},
	\end{equation}
	where $P$ and $M$ are shift and rotation generators, respectively, $ '+' $ sign corresponds to the negative curvature, and $'-'$ sign --- to the positive one. 
	At the level of measurements, in turn, it means that in a curved space, one can not measure distances using straight rods with arbitrary precision: the distance between two points can be measured either by extremals of nonzero curvature or by straight but non-geodesic lines.

	Another example of uncertainty relation with the presence of universal constant is Bohr-Rosenfeld relation in statistical physics, which connects uncertainties of measurements of thermodynamic functions. Bohr proposed a thought experiment in which the physical characteristics of the thermodynamic system are measured in different conditions. First, a system is encased in an adiabatic shell, so it is unable to exchange energy with the environment. Then it is possible to measure its energy with arbitrary precision, but one is unable to measure its temperature since it is not possible to bring it in contact with the thermometer without breaking the shell. On the other side, let us place the system in a heat bath. Then it becomes possible to measure its temperature precisely, but since such a system is exchanging energy with the heat bath, the precise measurement of its energy is no longer possible. 
	
	Rosenfeld gave this principle a quantitative form: 
	\begin{equation}\label{rosen}
	\Delta U \Delta \left(\frac{1}{T}\right) \geq k,
	\end{equation}
	where $U$ is the internal energy and $T$ is the temperature of the system. There are numerous ways to obtain these inequalities and other ones that connect other thermodynamic functions, as well as numerous interpretations of them. A detailed review of derivation and interpretation of thermodynamic uncertainty relations can be found, e.g., in \cite{Uffink1999}. 	The authors  propose their own derivation of these relations,  noting that in this version they are close to QM Heisenberg inequalities, since both are based on the general notion of "Local Heisenberg inequalities" established by L\'evy-Leblond \cite{LL1985}.
	
	However, the authors of this paper conclude that most other derivations of \eqref{rosen} are flawed and the Bohr's interpretation of it as a complementarity between states of a system in thermal isolation and in heat bath is disfavored. The reason of this conclusion is the lack of the equivalence between description of a system in terms of canonical ensemble (i.e. submerged in a heat bath) and microcanonical one (i.e. energetically isolated). Indeed, these descriptions are shown to be connected through Legendre transformation only for systems whose entropy is a concave function (i.e. systems without long-range interactions) \cite{1403.6608}. Moreover, it was pointed out by Mandelbrot \cite{mandelbrot} that these ensembles are related by \textit{Laplace} transformation, which is similar to the Fourier transformation connecting the position and momentum in QM as well as time and frequency in signal processing:
\begin{quote}
	One cannot fail to note the formal identity of this relationship with Heisenberg’s relationship, or its equivalent in communication: Gabor’s relation in the spectral analysis of signals. In fact, all three are one relation: they result from equality cases of Schwartz’s inequality, for dual variables. It does not matter, of course, whether the duality is Fourier, (quantum theory and spectral analysis) or Laplace (here).
	\end{quote}
	However, this situation only takes place in the thermodynamic limit, where Laplace and Legendre transformations "are intimately related to each other" \cite{0806.1147}.

	From the above consideration it can be seen that under some restrictions one can interpret  Boltzmann constant as playing the role in \eqref{rosen} similar to the one that $c$ plays in \eqref{gaborc} and $\hbar$ in \eqref{Heis} (although the area of applicability of \eqref{rosen} is much less wide than \eqref{gaborc} and \eqref{Heis}). Namely, it puts a limit on the possibility of a description of a system in "classical" (in this case, it means thermodynamical) terms. This limit is connected with the existence of more general theory, which describes this system in a different framework (e.g., statistical physics characterized by $k$, field theory characterized by $c$ or quantum theory characterized by $\hbar$). The $k\to 0$ limit of statistical physics is discussed, e.g., in \cite{1203.1479}, see also \cite{1701.01144} for another interpretation of this limit, which employs tropical geometry.
	
	Therefore we can not fully agree with the L\'evy-Leblond's characteristics of $k$ as "historical" constant: thermodynamics and statistical mechanics both have the status of modern physical theories (as classical and quantum mechanics), so $k$ is not just a unit conversion factor, but rather the constant which defines the area of applicability of thermodynamics. Namely, if an entropy of a system is comparable to $k$, one can not use a thermodynamic approach to describe it. 
	
	\section{General relativity and the status of \it{G}}\label{GR}
	
	Since in this paper we discussed the extension of the principle of relativity, some comments should be made about the extension that was proposed by Einstein: the principle of general relativity, which (as Einstein thought) unifies the notions of gravitation and inertia\cite{Lehmkuhl2014}. From the above considerations, a question naturally arises: is it possible to interpret the principle of general relativity in terms of relativity w.r.t. observational means? 
	
	Fock himself, who was a lifelong opponent of the notion of general relativity, had some doubts on that question. In one of his last papers \cite{Fock1973} he expressed his doubts as follows (we  present them here in full because the availability of the paper \cite{Fock1973} is quite limited):
	\begin{quotation}\label{fockrel}
		What kind of relativity do we want to accentuate by giving this name [special relativity] to Einstein's theory? It seems certain to us that it is precisely relativity w.r.t observational means, the latter being characterized by the inertial reference system to which they belong. But it seems much more difficult, if not impossible, to justify the name "general relativity" given to Einstein's theory of gravitation. Indeed, it is not possible to match each coordinate system with a laboratory one (containing means of observation), and even if it is assumed as possible, it must be admitted that the physical conditions in different laboratories are not the same, which means that in this case there is no relativity in the physical sense.
		
		However, the notion of relativity with respect to the means of observation is so general that it must be applicable to the theory of gravitation as well. It may be Einstein's equivalence principle that could allow this notion to be introduced more explicitly. To achieve this goal, we need to specify the nature of the means of observation and the area subject to their control. We should first define what is meant by local and global reference system. We could take the \emph{rep\'eres} (corresponding to a laboratory that moves without rotation along a geodesic) as a local system, and a privileged coordinate system, namely, adapted to the nature of the physical problem considered (e.g., harmonic coordinates for a system of bodies like Solar System)  as a global system. Analysis of the observation problem in Einstein's theory of gravity remains to be done.
	\end{quotation}
	As can be seen, Fock's belief in the universal character of PeRelOM has become so strong that he even sketched a way to justify the notion of general relativity through it: the notion that he staunchly denied throughout most of his career. However, he stressed that to perform such justification, one has to determine the observational means through which this relativity manifests itself. Indeed, in all examples discussed above, there were different ways to observe certain phenomena: position and momentum of a quantum system; transmission time and frequency band of a signal; energy and temperature of a thermodynamic system, etc. Therefore one needs to find such paired quantities in the Einsteinian gravitation theory in order to unveil two complementary descriptions of a physical system.
	
	L\'evy-Leblond proposed the following idea. Taking the trace of Einstein equations
	\begin{equation}
	\mathcal R = {8\pi G} \mathcal T, 
	\end{equation}
	where $\mathcal R$ and $\mathcal T$ are traces of the Ricci tensor and energy-momentum tensor, respectively, one can observe that "$[\mathcal R] = L^2$, $\mathcal [T] = ML^{-3}$, so that $[G] = LM^{-1}$, directly connecting mass and length (we have assumed that $c=1$ so that $[L]=[T]$)".\cite{LL1979}
	
	However, this consideration differs from the above ones by the fact that its validity depends on the number of spacetime dimensions, since in a spacetime with $n$ dimensions
	\begin{equation}
	[G_n] = L^{n-2} M, 
	\end{equation}
	if the convention $c=1$ is still adopted. Therefore, at $n=2$ the connection between mass and length cannot be established, and $G_2$ (being a type B constant) is not playing the role of "concept synthesizer". On the contrary, the universal constants $c, \hbar, R$, and $k$ (being type C constants) serve as manifestations of PeRelOM in a spacetime of arbitrary dimension.
	
	One can thus conclude that the Einsteinian gravitational theory should be interpreted not as a theory of relativity (see \eqref{fockrel}: "There is no relativity in physical sense"), but rather as a theory of specific interaction:\cite{LL1979}
	\begin{quote}
		From the precedingly described point of view, it is clear that general relativity a) is not a relativity, in that its geometrical interpretation and its invariance under coordinate transformation just "turn out", without any deep physical meaning; b) is not general, in that it is just the theory of gravitation, on the same level as the theory of electromagnetism or the theory (?) of strong interactions. In other terms, the constant $G$ here reverts to its type-B status.
	\end{quote}
		
	\section{Conclusion}
	In this paper we made a connestion between two methodological notions in the theoretical physics: the Fock's notion of relativity w.r.t. observational means and L\'evy-Leblond's notion of universal constant. It turned out that in different areas of physics the appearance of more fundamental theory brings out not only a certain universal constant, but also a limit on a simultaneous measurement of the characteristics of a physical system that correspond to preceding theory. We summarize all previous considerations in the following table:
	{\begin{table}[H]
		\caption{Universal constants in different theories and corresponding observational characteristics.}
		\centering
		\begin{adjustbox}{max width=\textwidth}
		\begin{tabular}{cccc}
			\toprule
			\textbf{"Classical" theory}	& \textbf{Classical concepts}	& \textbf{"Relativistic" theory} & \textbf{Universal relation} \\
			\midrule
			classical mechanics		& position \& momentum, \ldots			& quantum mechanics & $p=h/\lambda$\\
			action-at-a-distance		& time \& length			& field theory & $\lambda=c\tau$\\
			Euclidean geometry		& lengths \& angles			& Cayley-Klein geometry & $L =R \phi  $\\
			thermodynamics		& internal energy \& temperature, \ldots			& statistical physics & $U=kT$\\
			\bottomrule
		\end{tabular}
		\end{adjustbox}
	\end{table}
	It must be stressed that the very existence of these characteristics does not depend on a type of theory that describes the system: e.g., the notions of position and momentum are well-defined both in classical and quantum mechanics. They, therefore, remain in the theory even when the corresponding constant of type C vanishes in the "classical" limit (in contrast with constants of interaction (type B), whose vanishing means that corresponding interaction is absent). 
	
	Epistemologically, universal constants can be interpreted as quantities that allow us to connect the notions of our physical theories and our abilities to perceive the phenomena of reality. The human beings and their observational means are non-relativistic and non-quantum systems that are embedded in a continuous world with Euclidean geometry. As long as objects of their studies have the same characteristics, they are free to use "classical" models, which do not require $c$, $\hbar$, $R$, and $k$. The principle of relativity w.r.t. observational means, therefore, establishes the limits of such a description of reality.  We have no doubt that besides physics, there are other branches of human knowledge, in which similar relativity exists. V.~A.~Fock himself expressed such hopes in one of his last papers\cite{Fock1971r}:
	\begin{quote}
		As shown by the history of the development of science, general principles established in one field of knowledge could be valid in some other field. It seems that principle of relativity w.r.t. observational means has such general character. There lies its philosophical significance. 
	\end{quote}

	{\bf Acknowledgments.} {The author are grateful to I. Ado, A. Grib, M. Komarova, K. Pavlenko, S. Paston, T. Shumilov and A. Zabolotskiy for useful discussions, and to D. Kalinov and D. Lisachenko for the bibliographical assistance. The work is supported by RFBR Grant No.~20-01-00081.}


\begin{thebibliography}{10}
\newcommand{\enquote}[1]{``#1''}
\providecommand{\url}[1]{\texttt{#1}}
\providecommand{\urlprefix}{URL }
\expandafter\ifx\csname urlstyle\endcsname\relax
  \providecommand{\doi}[1]{doi:\discretionary{}{}{}#1}\else
  \providecommand{\doi}{doi:\discretionary{}{}{}\begingroup
  \urlstyle{rm}\Url}\fi
\providecommand{\eprint}[1]{\href{http://arxiv.org/abs/#1}{\texttt{#1}}}

\bibitem{2005.08196}
A.~Sheykin, S.~Manida, \enquote{Universal Constants and Natural Systems of
  Units in a Spacetime of Arbitrary Dimension},
  \href{http://dx.doi.org/10.3390/universe6100166}{\emph{Universe}},
  \textbf{6}: 10, \eprint{2005.08196}.

\bibitem{EPR}
A.~Einstein, B.~Podolsky, N.~Rosen, \enquote{Can Quantum-Mechanical Description
  of Physical Reality Be Considered Complete?},
  \href{http://dx.doi.org/10.1103/PhysRev.47.777}{\emph{Phys. Rev.}},
  \textbf{47} (1935), 777--780.

\bibitem{Fock1936}
V.~A. Fock, Introductory note to Russian translation of \cite{EPR},
  \href{http://dx.doi.org/10.3367/UFNr.0016.193604b.0436}{\emph{Sov. Usp.
  Phys.}}, \textbf{16}: 4 (1936), 436--457.

\bibitem{graham}
L.~R. Graham, \enquote{Science, Philosophy, and Human Behavior in the Soviet
  Union}, Columbia University Press, 1989.

\bibitem{Fock1938}
V.~A. Fock, \enquote{On discussion on questions of physics (in Russian)},
  \emph{Pod znamenem marksizma}, \textbf{1} (1938), 149--159.

\bibitem{Fock1949}
V.~A. Fock, \enquote{Main laws of physics in the light of dialectic
  matherialism (in Russian)}, \emph{Vestn. LGU}, \textbf{4} (1949), 34--47.

\bibitem{Bohr1948}
N.~Bohr, \enquote{On the notions of causality and complementarity},
  \href{http://dx.doi.org/10.1111/j.1746-8361.1948.tb00703.x}{\emph{Dialectica}},
  \textbf{2}: 3‐4 (1948), 312--319.

\bibitem{Fock1955}
V.~A. Fock, \enquote{A criticism of Bohr's quantum-mechanical concepts},
  \href{http://dx.doi.org/10.1007/BF01687207}{\emph{Chechosl. J. Phys.}},
  \textbf{5}: 4 (1955), 448--448.

\bibitem{martinez2019}
J.-P. Martinez, \enquote{Beyond Ideology: Epistemological Foundations of
  Vladimir Fock's approach to Quantum Theory},
  \href{http://dx.doi.org/https://doi.org/10.1002/bewi.201900008}{\emph{Ber.
  Wissenschaft.}}, \textbf{42}: 4 (2019), 400--423.

\bibitem{Fock1957}
V.~A. Fock, \enquote{On the interpretation of quantum mechanics},
  \href{http://dx.doi.org/10.1007/BF01946586}{\emph{Czechosl. J. Phys.}},
  \textbf{7}: 6 (1957), 643--656.

\bibitem{bondi1968}
H.~Bondi, \enquote{Cosmology}, Cambridge monographs on physics, Cambridge
  University Press, 2nd edn., 1968.

\bibitem{yaglom1979}
I.~M. Yaglom, \enquote{A simple non-Euclidean geometry and its physical basis:
  an elementary account of Galilean geometry and the Galilean principle of
  relativity}, Heidelberg science library, Springer, 1979.

\bibitem{einstein1920}
A.~Einstein, \enquote{Relativity: The Special and the General Theory (100th
  Anniversary Edition)}, Princeton University Press, 100th anniversary edn.,
  2015.

\bibitem{chiatti2014}
L.~Chiatti, I.~Licata, \enquote{Relativity with Respect to Measurement:
  Collapse and Quantum Events from Fock to Cramer},
  \href{http://dx.doi.org/10.3390/systems2040576}{\emph{Systems}}, \textbf{2}:
  4 (2014), 576--589.

\bibitem{quant-ph/9609002}
C. Rovelli, \enquote{Relational quantum mechanics},
  \href{http://dx.doi.org/10.1007/bf02302261}{\emph{Int. J. Theor. Phys.}}, \textbf{35}: 8 (1996), 1637–1678,
  \eprint{quant-ph/9609002}.

\bibitem{Fock1971}
V.~A. Fock, \enquote{Quantum physics and philosophical problems},
  \href{http://dx.doi.org/10.1007/BF00708579}{\emph{Found. Phys.}}, \textbf{1}:
  4 (1971), 293--306.

\bibitem{Fock1973}
V.~A. Fock, \enquote{Le principe de relatlvite par rapport aux moyens
  d'observation}, in \emph{Symposia Mathematica}, vol.~12, 327--335, Istituto
  Nazionale di Alta Matematica, Bologna, 1973.

\bibitem{barrow2003constants}
J.~D. Barrow, \enquote{The Constants of Nature: From Alpha to Omega}, Vintage,
  2003.

\bibitem{LL1977}
J.-M. L{\'e}vy-Leblond, \enquote{On the conceptual nature of the physical
  constants}, \href{http://dx.doi.org/10.1007/BF02748049}{\emph{Riv. Nuovo
  Cim.}}, \textbf{7}: 2 (1977), 187--214.

\bibitem{LL1981}
J.-M. L\'evy-Leblond, \enquote{Classical apples and quantum potatoes},
  \href{http://dx.doi.org/10.1088/0143-0807/2/1/007}{\emph{Eur. J. Phys.}},
  \textbf{2}: 1 (1981), 44--47.

\bibitem{gabor1946}
D.~Gabor, \enquote{Theory of communication. Part 1: The analysis of
  information}, \href{http://dx.doi.org/10.1049/ji-3-2.1946.0074}{\emph{J.
  Inst. Electr. Eng. 3.}}, \textbf{93}: 26 (1946), 429--441.

\bibitem{Grochenig2001}
K.~Gr{\"o}chenig, Foundations of Time-Frequency Analysis, ch. 3: \href{10.1007/978-1-4612-0003-1_3}{Time-Frequency Analysis and the Uncertainty Principle}, pp.
  21--36, Birkh{\"a}user Boston, Boston, MA, 2001.

\bibitem{1208.4611}
J.~N. Oppenheim, M.~O. Magnasco, \enquote{Human Time-Frequency Acuity Beats the
  Fourier Uncertainty Principle},
  \href{http://dx.doi.org/10.1103/physrevlett.110.044301}{\emph{Phys. Rev.
  Lett.}}, \textbf{110}: 4, \eprint{1208.4611}.

\bibitem{Uffink1999}
J.~Uffink, J.~van Lith, \enquote{Thermodynamic Uncertainty Relations},
  \href{http://dx.doi.org/10.1023/A:1018811305766}{\emph{Found. Phys.}},
  \textbf{29}: 5 (1999), 655--692.

\bibitem{LL1985}
J.-M. L\'evy-Leblond, \enquote{Local Heisenberg inequalities},
  \href{http://dx.doi.org/https://doi.org/10.1016/0375-9601(85)90367-6}{\emph{Phys.
  Lett. A}}, \textbf{111}: 7 (1985), 353 -- 355.

\bibitem{1403.6608}
H.~Touchette, \enquote{Equivalence and Nonequivalence of Ensembles:
  Thermodynamic, Macrostate, and Measure Levels},
  \href{http://dx.doi.org/10.1007/s10955-015-1212-2}{\emph{J. Stat. Phys.}},
  \textbf{159}: 5 (2015), 987–1016, \eprint{1403.6608}.

\bibitem{mandelbrot}
B.~{Mandelbrot}, \enquote{An outline of a purely phenomenological theory of
  statistical thermodynamics--I: Canonical ensembles},
  \href{http://dx.doi.org/10.1109/TIT.1956.1056804}{\emph{IRE Trans. Inf.
  Theory}}, \textbf{2}: 3 (1956), 190--203.

\bibitem{0806.1147}
R.~K.~P. Zia, E.~F. Redish, S.~R. McKay, \enquote{Making sense of the Legendre
  transform}, \href{http://dx.doi.org/10.1119/1.3119512}{\emph{Am. J. Phys.}},
  \textbf{77}: 7 (2009), 614–622, \eprint{0806.1147}.

\bibitem{1203.1479}
L.~Velazquez~Abad, \enquote{Principles of classical statistical mechanics: A
  perspective from the notion of complementarity},
  \href{http://dx.doi.org/10.1016/j.aop.2012.03.002}{\emph{Annals of Physics}},
  \textbf{327}: 6 (2012), 1682–1693, \eprint{1203.1479}.

\bibitem{1701.01144}
M.~Angelelli, \enquote{Tropical limit and a micro-macro correspondence in
  statistical physics},
  \href{http://dx.doi.org/10.1088/1751-8121/aa863b}{\emph{J. Phys. A:
  Math. Theor.}}, \textbf{50}: 41 (2017), 415202,
  \eprint{1701.01144}.

\bibitem{Lehmkuhl2014}
D.~Lehmkuhl, \enquote{{Why Einstein did not believe that general relativity
  geometrizes gravity}},
  \href{http://dx.doi.org/10.1016/j.shpsb.2013.08.002}{\emph{Stud. Hist. Phil.
  Sci. B}}, \textbf{46} (2014), 316--326.

\bibitem{LL1979}
J.-M. L\'evy-Leblond, \enquote{The importance of being (a) constant}, in
  \href{https://inis.iaea.org/search/search.aspx?orig_q=RN:11503317}{\emph{Problems in the foundations of physics}}, edited by G.~Toraldo~di
  Francia, 237--263, North-Holland, Amsterdam, 1979.

\bibitem{Fock1971r}
V.~A. Fock, \enquote{The principle of relativity with respect to the
  observational means in the modern physics (in Russian)}, \emph{Vestn. AN
  SSSR}, \textbf{4} (1971), 8--12.

\end{thebibliography}

\end{document}